\documentclass[10pt,
floatfix, 
aps, 
pra, 
amsmath,
twocolumn,
superscriptaddress
]{revtex4-2}

\pdfoutput=1

\usepackage{graphicx}
\usepackage{amsmath}
\usepackage{amssymb}
\usepackage{bbm}
\usepackage[colorlinks=true, allcolors=blue]{hyperref}
\usepackage{soul}
\usepackage{xcolor}
\usepackage{bbold}
\usepackage{physics}
\usepackage{mathtools}

\begin{document}

\title{Evidence for a low-gap component in compressively stressed niobium thin films}

\author{J. T. Paustian}
\affiliation{Department of Physics, Syracuse University, Syracuse, NY 13244, USA}
\affiliation{Institute for Quantum and Information Sciences,
Syracuse University, Syracuse, NY 13244, USA}

\author{Maciej Olszewski}
\affiliation{Department of Physics, Cornell University, Ithaca, NY 14853, USA}
\affiliation{School of Applied and Engineering Physics, Cornell University, Ithaca, NY 14853, USA}

\author{Kiichi Okubo}
\affiliation{Department of Physics, University of Wisconsin--Madison, Madison, WI 53706, USA}

\author{Lingda Kong}
\affiliation{Department of Physics, Cornell University, Ithaca, NY 14853, USA}
\affiliation{School of Applied and Engineering Physics, Cornell University, Ithaca, NY 14853, USA}

\author{B. L. T. Plourde}
\affiliation{Department of Physics, Syracuse University, Syracuse, NY 13244, USA}
\affiliation{Department of Physics, University of Wisconsin--Madison, Madison, WI 53706, USA}

\author{Ethan~G.~Arnault}
\affiliation{Department of Physics, Syracuse University, Syracuse, NY 13244, USA}
\affiliation{Institute for Quantum and Information Sciences,
Syracuse University, Syracuse, NY 13244, USA}
\affiliation{
Department of Electrical Engineering and Computer Science, Syracuse University, Syracuse, NY 13244, USA
}%

\author{Caleb W. Fink}
\affiliation{Department of Physics, Syracuse University, Syracuse, NY 13244, USA}
\affiliation{Institute for Quantum and Information Sciences,
Syracuse University, Syracuse, NY 13244, USA}

\author{Valla Fatemi}
\affiliation{School of Applied and Engineering Physics, Cornell University, Ithaca, NY 14853, USA}

\author{Ivan V. Pechenezhskiy}
\email{ivpechen@syr.edu}
\affiliation{Department of Physics, Syracuse University, Syracuse, NY 13244, USA}
\affiliation{Institute for Quantum and Information Sciences,
Syracuse University, Syracuse, NY 13244, USA}

\date{\today}

\begin{abstract}
We investigate the microwave response of sputtered Nb films as a function of biaxial film stress. Although transport measurements show nearly identical superconducting transition temperatures ($9.0\pm 0.2$\,K) and residual resistance ratios across the stress series, microwave resonators fabricated from highly compressive films exhibit enhanced temperature-dependent loss and larger frequency shifts. A conventional single-phase Mattis--Bardeen description yields an effective critical temperature of $2.4\pm0.2$\,K, approximately a factor of four below the value obtained from transport measurements. We show that the data are instead consistent with the presence of an additional low-gap component whose contribution increases with compressive stress. These results show that conventional transport measurements can overlook stress-dependent low-energy excitations that strongly influence microwave performance.
\end{abstract}

\maketitle

\section{Introduction}

Superconducting thin films underpin a wide range of quantum devices and detectors, including superconducting qubits~\cite{Siddiqi2021}, microwave resonators~\cite{McRae2020}, kinetic-inductance detectors~\cite{Mazin2004}, transition-edge sensors~\cite{Irwin2005}, superconducting nanowire single-photon detectors~\cite{Oripov_2023}, and qubit-based cryogenic particle detectors~\cite{Fink_2024}. In each of these systems, device performance is ultimately determined by the low-energy electrodynamic response of the superconducting film, which defines microwave dissipation, kinetic inductance, and quasiparticle dynamics. The corresponding low-energy excitations limit coherence times in superconducting qubits~\cite{Siddiqi2021}, degrade the energy resolution of cryogenic detectors~\cite{Anthony-Petersen2025}, and produce excess noise in superconducting microwave circuits~\cite{Gao2008, Krantz2019}.

In practice, the superconducting film quality is commonly evaluated using metrics such as the superconducting transition temperature $T_c$ and the residual resistance ratio (RRR), and by probing the crystal structure. However, these measurements are more sensitive to the equilibrium or highest-$T_c$ superconducting properties than to the low-energy excitations relevant to microwave operation.
As a result, films that appear nearly identical by conventional materials characterization can in principle exhibit drastically different performance in high-quality-factor superconducting microwave devices.

Among the many properties that characterize superconducting films, intrinsic film stress is known to modify microstructure, defect density, and crystalline phase stability~\cite{Abadias2018}. 
These relationships suggest the possibility that stress may also influence the superconductor's low-energy electrodynamic response. Such low-energy excitations have attracted considerable attention in superconducting device physics because they are believed to contribute to excess microwave dissipation and frequency noise in superconducting resonators and qubits~\cite{Gao2008, McRae2020, Siddiqi2021}, nonequilibrium quasiparticle generation~\cite{Anthony-Petersen2024}, and low-energy event backgrounds in cryogenic particle detectors~\cite{Romani2024, LEE}.

While the superconducting properties of sputtered Nb thin films are well characterized~\cite{Wolf1976, Minhaj1994, Lacquaniti1997}, the influence of sputter-deposition conditions and film structure on superconducting microwave properties merits reexamination. Here, we systematically investigate the microwave response of coplanar waveguide (CPW) resonators fabricated from Nb films spanning a nearly 1-GPa range of intrinsic biaxial stress. Although all films exhibit nearly identical transport properties, including superconducting transition temperatures $T_c$ near 9\,K, their microwave response changes dramatically with stress. Temperature-dependent loss and frequency-shift measurements cannot be reconciled with a conventional single-phase response, but a two-component model consistently accounts for them via an additional low-gap contribution with a critical temperature of about 2.4\,K. As compressive stress increases, the contribution of this component becomes evident in the temperature-dependent response; for the most compressive film, the internal quality factors are degraded even at the lowest measured temperature.

\begin{figure*}[t]
    \centering
    \includegraphics[width=\linewidth]{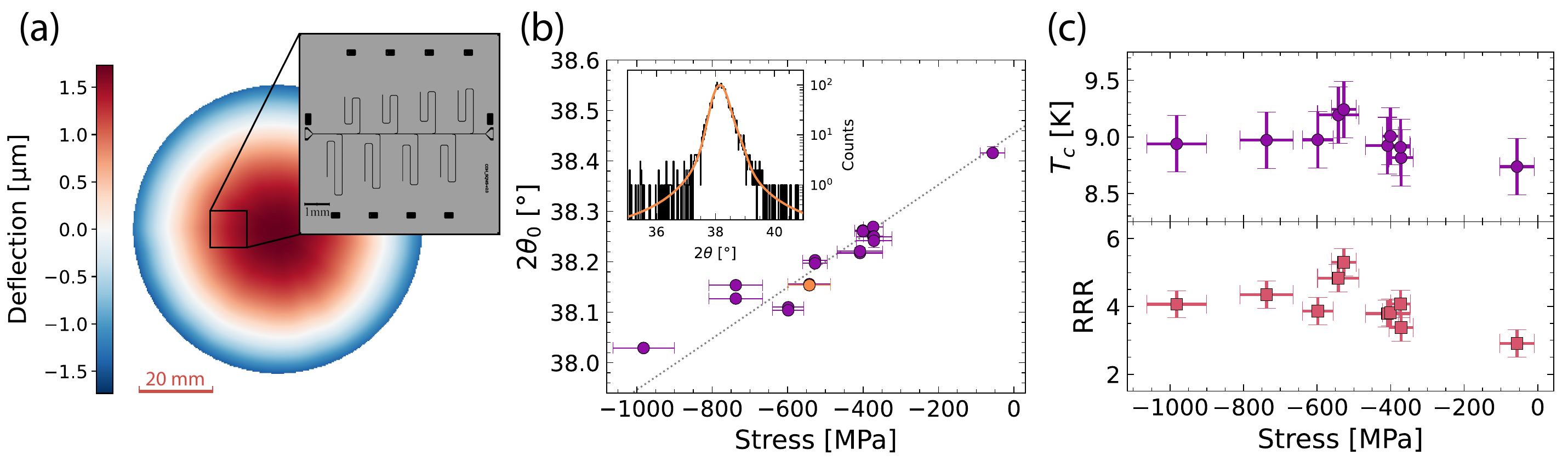}
    \caption{\textbf{Structural and transport characterization of stressed Nb films.} (a)~Wafer-deflection map corresponding to the orange data point in panel (b), obtained from stylus profilometry of the wafer before and after Nb deposition; the resulting surface curvature is used to extract the in-plane biaxial film stress. Positive and negative deflections indicate the spatial variation of the wafer bow. The inset shows a device layout with CPW resonators coupled to a common feedline and patterned from the stressed Nb film. 
    (b)~The bcc-Nb(110) peak center $2\theta_0$ as a function of film stress. The peak center is determined by fitting the diffraction peak, as shown in the inset for the XRD spectrum corresponding to the orange data point. Increased compressive stress shifts the Nb(110) peak to a lower angle, indicating an expansion of the out-of-plane lattice spacing, consistent with in-plane compressive film strain. The dotted line is a linear fit to the data. The observed monotonic stress-dependent peak shift provides an independent check of the profilometry-based stress measurement.
    (c)~Four-probe dc transport properties of Nb films versus film stress. The upper panel shows the measured superconducting transition temperature $T_c$, and the lower panel shows the residual resistance ratio (RRR). The vertical error bars represent the variation across samples fabricated under similar conditions. Across the entire stress range, the films retain $T_c$ values near 9.0\,K and comparable RRR values, indicating that intrinsic film stress does not strongly affect the dc superconducting transition or normal-state transport.}
    \label{fig:Fig1}
\end{figure*}

\section{Experimental methods}

\subsection{Film stress characterization}

Niobium films of varying in-plane biaxial stress were sputter-deposited under various conditions. We varied the target-to-substrate distance (sometimes referred to as the ``throw distance''), the argon gas pressure, and the deposition time, which sets the desired film thickness, as summarized in Appendix~\ref{appendix:fab}, Table~\ref{tab:sputter}. The intrinsic stress of the film is primarily controlled by the momentum of the Nb atoms on impact with the substrate, which is determined by the Ar gas pressure and the throw distance, as both influence the scattering of the Nb atoms in flight~\cite{Windischmann1991}. The deposition time, and thus the film thickness, had only a mild effect on the measured stress.

Throughout this work, compressive film stress is taken as negative and tensile film stress as positive. Film stress characterization was performed at room temperature using a stylus profilometer (Appendix~\ref{appendix:profilometery}) on a separate set of wafers sputtered under the same conditions as those used for resonator fabrication. A deflection map for the $-542\,\mathrm{MPa}$ film is shown in Fig.~\ref{fig:Fig1}(a). Figure~\ref{fig:Fig1}(b) shows that the profilometry-derived film stress is strongly correlated with the Nb(110) peak position in X-ray diffraction (XRD) measurements (Appendix~\ref{appendix:XRD}).

\subsection{Transport characterization}

We performed four-probe resistance measurements on the same films used for profilometry using a Quantum Design DynaCool physical property measurement system (PPMS) over the temperature range from room temperature to 1.8\,K. As shown in Fig.~\ref{fig:Fig1}(c), the films have an~average $T_c$ of $9.0\pm0.2$\,K and an average RRR of $4.0\pm0.6$. We find no apparent correlation between film stress and RRR or $T_c$, nor between microwave losses and RRR or $T_c$, and attribute the observed minor variations in RRR and $T_c$ to possible variations in trapped argon concentration~\cite{Benvenuti1995}.

\subsection{Microwave measurements}

To characterize microwave loss in the films spanning a range of biaxial stresses, we fabricated test devices with eight quarter-wave CPW resonators on each chip spanning the 4.2--6.3\,GHz frequency range. We used the standardized layout of Ref.~\cite{Kopas22} shown in Fig.~\ref{fig:Fig1}(a) with a~6\,$\mu$m CPW inner trace width and a 3\,$\mu$m CPW gap. The device fabrication process is described in detail in Appendix~\ref{appendix:fab}. The microwave loss of the CPW resonators was characterized in an~adiabatic demagnetization refrigerator setup (Appendix~\ref{appendix:setup}) by measuring the microwave transmission through the on-chip feedline to which all resonators are coupled. To extract the resonant frequency $f$ and the internal quality factor~$Q_i$ for each resonator, we fit the inverse of the transmission near each resonance to a circle following Ref.~\cite{Megrant_2012}. We measured the resonators as a function of both microwave power and temperature. We first performed power sweeps at 100\,mK to characterize saturation of the two-level system (TLS) loss. The extracted saturation photon numbers were then used to select the microwave drive powers for the temperature-dependent measurements, which were performed well within the TLS-saturated regime, so that resonant TLS loss is substantially suppressed during the temperature-dependent measurements (Appendices~\ref{appendix:photon_number} and~\ref{appendix:power}). Temperature sweeps were performed from 100\,mK to 3.0\,K for all devices; however, we restricted the data analysis to temperatures below 1\,K to remain below the $T_c$ of the aluminum wire bonds. 

\begin{figure}
    \centering
    \includegraphics[width=\linewidth]{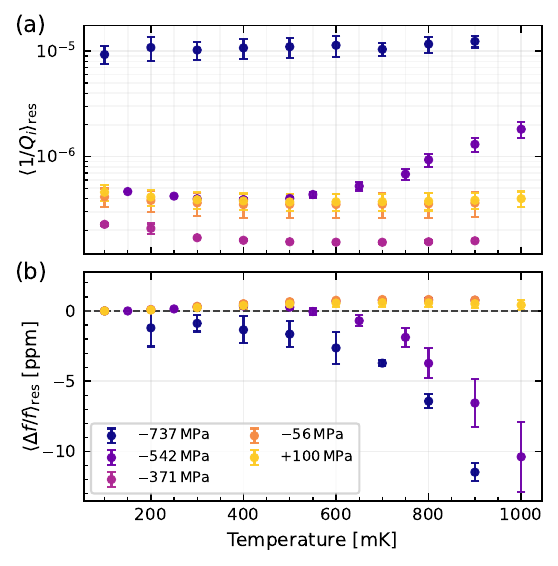}
    \caption{
    \textbf{Average microwave response of the CPW resonators.} For each sample, the inverse internal quality factor $1/Q_i$ and the relative frequency shift $\Delta f/f$, referenced to the lowest temperature, are averaged over the resonators, yielding $\langle 1/Q_i\rangle_{\rm res}$ and  $\langle \Delta f/f\rangle_{\rm res}$. Error bars show the standard deviation across resonators from the same sample. Highly compressive films exhibit greater loss and larger temperature-dependent frequency shifts, indicating that the microwave response is sensitive to intrinsic film stress.}
    \label{fig:empirical_loss}
\end{figure}

\section{Experimental observations}

Although the dc transport measurements show little dependence on film stress, the microwave response varies strongly over the same stress range. Because the film stress is largely uniform across each wafer, we first analyze the microwave response by averaging the responses of resonators from the same sample for clarity. The average inverse internal quality factor $\langle 1/Q_i\rangle_{\rm res}$ and the average referenced frequency shift $\langle \Delta f/f\rangle_{\rm res}$ are shown in Fig.~\ref{fig:empirical_loss}.

The average microwave loss $\langle 1/Q_i\rangle_{\rm res}$ shows a clear dependence on the film stress. Films with in-plane biaxial stress more compressive than approximately $-500$\,MPa exhibit enhanced loss, with the $-737$-MPa film exhibiting the largest loss, of order $10^{-5}$. In contrast, films with stresses less compressive than approximately $-400$\,MPa exhibit lower losses and weaker temperature dependence over the same temperature range. The frequency-shift data show a complementary stress-dependent response: highly compressive films exhibit larger temperature-dependent shifts, whereas less compressive ones show a reduced response. Among the measured films, the sample with room-temperature stress of \mbox{$-371$\,MPa} yielded the highest-quality resonators (see also Fig.~\ref{fig:power_fits}).

\begin{figure*}
    \centering
    \includegraphics[width=\linewidth]{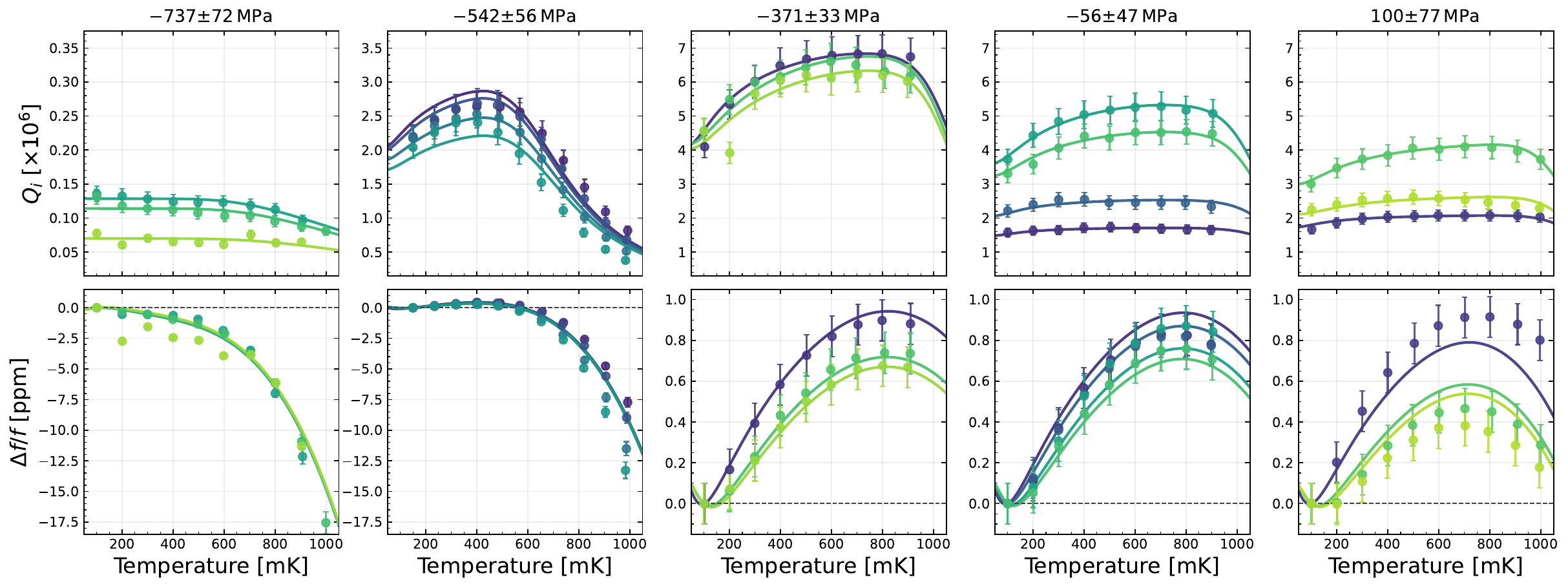}
    \caption{\textbf{Temperature-dependent microwave response at various film stresses.} Internal quality factors $Q_i$ (top row) and referenced frequency shifts $\Delta f/f$ (bottom row) for Nb CPW resonators fabricated from films with in-plane biaxial stress as indicated on top of each column. Lighter-colored data points correspond to higher-frequency resonators, while solid curves of the same color show fits to the two-component model. The loss data determine the effective secondary-component critical temperature $T_c^\textrm{2nd} = 2.4\pm0.2$\,K, while the frequency response is fit conditionally using the already determined~$T_c^\textrm{2nd}$. Highly compressive films show enhanced temperature-dependent loss and larger reactive shifts, consistent with an increased contribution from the secondary component.
 }
    \label{fig:Qidf_Model}
\end{figure*}

\section{Microwave response analysis}

\subsection{Motivation for a two-component model}

The sample-averaged measurements in Fig.~\ref{fig:empirical_loss} establish a clear correlation between film stress and microwave response, since averaging emphasizes contributions common to resonators from the same film while suppressing resonator-to-resonator variability. Highly compressive films exhibit both enhanced loss and larger resonator frequency shifts with temperature, while less compressively stressed films show comparatively weak temperature dependence in both quantities.

At first glance, these trends might be attributed to thermally generated quasiparticles within a conventional single-phase bcc-Nb film. However, this interpretation leads to a contradiction. The transport measurements presented above show that all films remain superconducting near the expected transition temperature of bcc-Nb, with only a modest variation across the stress series. In contrast, fitting the temperature-dependent microwave loss using a conventional Mattis--Bardeen quasiparticle model~\cite{Mattis-Bardeen1958} yields a critical temperature substantially below the transport transition temperature. A homogeneous reduction of the superconducting gap throughout the film cannot explain the discrepancy because a~correspondingly large suppression of the transport transition temperature would be expected, contrary to observation. Instead, the measurements suggest an additional contribution to the microwave response that strongly affects the complex conductivity at microwave frequencies, but exerts only a limited influence on dc transport.

The systematic dependence on stress provides an additional clue. The temperature-dependent microwave anomalies become progressively stronger with increasing compressive stress, indicating that the effect likely originates from a stress-dependent modification of the film microstructure rather than from measurement artifacts or resonator-specific defects. The simplest interpretation is therefore that the films contain a dominant bcc-Nb superconducting phase, together with an additional contribution from a~secondary, lower-gap component, whose effective contribution increases with compressive stress.

\subsection{Two-component model}
We interpret the data using a two-component model. We phenomenologically represent the effective film conductivity $\sigma^\mathrm{eff}$ as a weighted sum of the conductivities of the dominant bcc-Nb component and a secondary low-gap component:
\begin{align}
    \sigma^\mathrm{eff}(T,\sigma_\textrm{stress}) =
    [1-p(\sigma_\textrm{stress})]&\sigma^\mathrm{bcc}(T) \nonumber\\
    + p(\sigma_\textrm{stress})&\sigma^{\mathrm{2nd}}(T),
    \label{eq:sigma_eff_schematic}
\end{align}
where $T$ is the temperature, $\sigma_\textrm{stress}$ is the film stress, $\sigma^\mathrm{bcc}(T)$ and  $\sigma^{\mathrm{2nd}}(T)$ are the temperature-dependent complex conductivities of the bcc-Nb and secondary components, and $p(\sigma_\textrm{stress})$ is an~effective stress-dependent fraction of the secondary component. The real part of the effective conductivity contributes to the total loss, while the imaginary part causes a frequency shift via a~change in kinetic inductance. Because several factors can prevent reliable estimation of the secondary component fraction, we do not extract $p$ directly; instead, we represent its dissipative and reactive contributions below as independent effective amplitudes.

For resonator $j$ from sample $s$, the loss model is written as
\begin{align}
\frac{1}{Q_{i;s,j}(T)} = \delta_{0;s,j} &+ A^\mathrm{TLS}_s\, g^\mathrm{TLS}(f_{s,j},T) \nonumber\\
& + A_s^\mathrm{2nd}\, g^\mathrm{QP}(f_{s,j},T;T_c^\mathrm{2nd}),
\label{eq:minimal_loss_model}
\end{align}
where $\delta_{0;s,j}$ is a temperature-independent resonator-specific loss floor that captures residual losses not explicitly included in the model, such as fabrication-dependent loss, packaging loss, and radiative loss. We discuss the two temperature-dependent terms below.

The first temperature-dependent contribution,
\begin{equation}
g^{\rm TLS}(f,T) = \tanh\left(\frac{h f}{2k_BT}\right),
\end{equation}
describes the temperature dependence of TLS absorption at the resonator frequency $f_{s,j}$, where $h$ is Planck's constant and $k_B$ is Boltzmann's constant~\cite{McRae2020}. Since the temperature sweeps were performed at photon numbers well above the fitted TLS saturation scale (Appendix~\ref{appendix:power}), the photon-number saturation factor is absorbed into the fitted amplitude $A^{\rm TLS}_s$. Consequently, $A^{\rm TLS}_s$ represents the residual power-saturated TLS contribution for sample $s$ in the temperature sweeps.

The quasiparticle basis function $g_\mathrm{QP}$ is computed from the complex Mattis--Bardeen conductivity $\sigma=\sigma_1-i\sigma_2$:
\begin{equation}
g^\mathrm{QP}(f,T;T_c) = \frac{\sigma_1(f,T;T_c)}{\sigma_2(f,T;T_c)}.
\end{equation}
The explicit approximate Mattis--Bardeen forms used to compute $g_{\rm QP}$ are given in Appendix~\ref{appendix:MB}. This function therefore provides the temperature-dependent shape of the microwave loss expected from thermally excited quasiparticles in a superconductor with transition temperature~$T_c$. 

The corresponding frequency shift $\Delta f_{s,j}(T) = f_{s,j}(T) - f_{s,j}(T_{\rm ref})$ of resonator~$j$ from sample~$s$ referenced to $T_\mathrm{ref}$ is modeled as
\begin{equation}
\frac{\Delta f_{s,j}(T)}{f_{s,j}(T_{\rm ref})}=
B^{\rm TLS}_s \Delta h^{\rm TLS}(f_{s,j},T) + B^{\rm 2nd}_s k^{\rm QP}(f_{s,j},T;T_c^{\rm 2nd}),
\label{eq:freq_shift}
\end{equation}
where the first term represents the reactive response of the TLS bath, and the second term accounts for the kinetic inductance contribution. The TLS bath response term, weighted by amplitude $B^{\rm TLS}_s$, is defined as
\begin{equation}
\Delta h^{\rm TLS}(f,T) = h^{\rm TLS}(f,T) - h^{\rm TLS}(f,T_{\rm ref})
\end{equation}
with $T_{\rm ref} = 100$\,mK and
\begin{equation}
h^{\rm TLS}(f,T) = \frac{1}{\pi}
\left[\Re\psi\!\left(\frac12+\frac{hf}{2\pi i k_BT}\right) 
-\ln\!\left(\frac{hf}{2\pi k_BT}\right)\right],\nonumber
\end{equation}
where $\psi$ is the digamma function~\cite{Gao2008,McRae2020}. Because the kinetic inductance $L_k$ is inversely proportional to the imaginary conductivity $\sigma_2$, we define the kinetic-inductance basis function in Eq.~\eqref{eq:freq_shift} as
\begin{equation}
k^{\rm QP}(f,T;T_c)=\frac{1}{2}\left[\frac{\sigma_2(f,T;T_c)}{\sigma_2(f,T_{\rm ref};T_c)}-1\right].
\end{equation}
To leading order, this kinetic-inductance basis function, weighted by the kinetic-inductance fraction, determines the relative frequency shift $\Delta f/f\simeq[L_k/(L_g+L_k)]k^{\rm QP}$, since a small change $\Delta L_k$ in kinetic inductance produces a relative frequency shift
$\Delta f/f\simeq-\Delta L_k/[2(L_g+L_k)]$. Here, $L_g$ is the geometric inductance and $L_k$ is evaluated at $T_\mathrm{ref}$. As the quasiparticle density increases, the superfluid density and reactive superfluid conductivity~$\sigma_2$ decrease, increasing the kinetic inductance and producing the observed redshift of the resonator frequencies. In the two-component model, the sample-dependent amplitude $B^{\rm 2nd}_s$ absorbs the kinetic-inductance fraction $L_k/(L_g+L_k)$. Because the model does not specify the geometry or distribution of the secondary component, $B^{\rm 2nd}_s$ should not be interpreted as a direct measurement of the kinetic-inductance fraction or the volume fraction of the secondary low-gap component, but rather as the effective microwave contribution of the secondary component in the reactive response of the resonators.

\subsection{Fit to the two-component model}

Figure~\ref{fig:Qidf_Model} shows internal quality factors $Q_i$ and relative frequency shifts $\Delta f/f$ as functions of temperature~$T$ for the five film stresses investigated. Except for the $-737$\,MPa sample, all devices show an initial increase in $Q_i$ and a blueshift in resonance frequency, consistent with residual thermal saturation of TLS. At higher temperatures, all devices exhibit a decrease in $Q_i$ and a redshift of the resonance frequency, consistent with an increase in the thermal quasiparticle population. The absence of a visible TLS signature in the $-737$\,MPa sample is consistent with its elevated power-independent loss (see Fig.~\ref{fig:power_fits}), which masks the residual TLS contribution.

We fit the two-component model presented above sequentially to the dissipative and reactive responses. We first fit the temperature-dependent loss of all resonators simultaneously using Eq.~\eqref{eq:minimal_loss_model}, with a single common secondary-component transition temperature~$T_c^{\rm 2nd}$, sample-dependent amplitudes $A_s^{\rm TLS}$ and $A_s^{\rm 2nd}$, and resonator-dependent background losses $\delta_{0;s,j}$. The loss data therefore determine $T_c^{\rm 2nd}$ and its uncertainty (the one-sigma interval from the likelihood profile). We then fit the frequency-shift data using Eq.~\eqref{eq:freq_shift}, fixing $T_c^{\rm 2nd}$ to the value inferred from the loss fit and allowing the sample-dependent amplitudes $B_s^{\rm TLS}$ and $B_s^{\rm 2nd}$ to vary.

The thermally activated quasiparticle loss gives direct sensitivity to the superconducting gap of the secondary component. Fitting the loss data yields $T_c^\textrm{2nd} = 2.4\pm0.2$\,K. If this temperature were instead interpreted as the transition temperature of a homogeneous single-phase Nb film, it would be incompatible with the $T_c = 9.0\pm 0.2$\,K measured by dc transport. The frequency-shift data are then fit conditionally using the loss-derived $T_c^\textrm{2nd}$, and the resulting resonator-level fits are shown in Fig.~\ref{fig:Qidf_Model}. A single fixed value of $T_c^\textrm{2nd}$ is sufficient to capture temperature-dependent losses and frequency shifts across the entire sample set over the explored temperature range, showing that a common low-gap component consistently describes the dissipative and reactive responses. For the weakly stressed films, the quasiparticle contribution is small over the measured temperature range. Measurements extended to sufficiently high temperatures without the additional loss from the Al stitching bonds (Appendix~\ref{appendix:setup}) would therefore be expected to approach the response of the dominant bcc-Nb component with $T_c\approx T_c^\textrm{bcc}$.

\begin{figure}
    \centering
    \includegraphics[width=\linewidth]{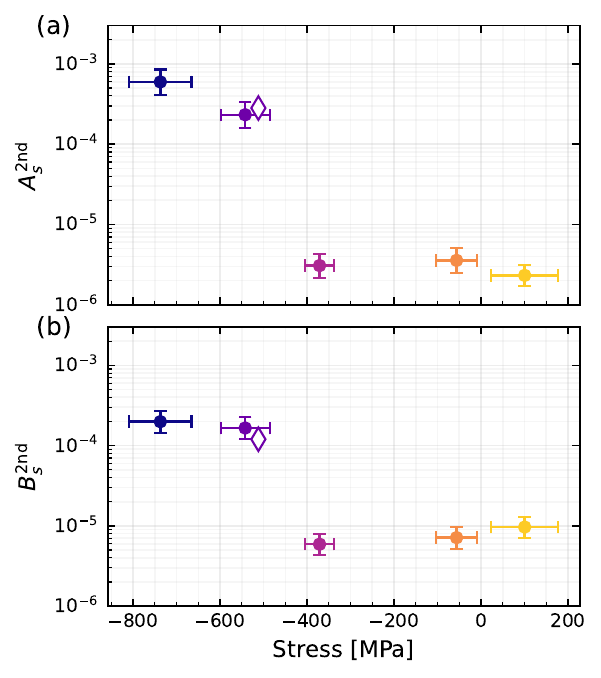}
    \caption{
\textbf{Stress dependence of the effective secondary-component amplitudes in the microwave response of the films.} (a)~Effective dissipative amplitude $A_s^{\rm 2nd}$ extracted from the resonator-level loss fits. (b)~Effective kinetic-inductance amplitude $B_s^{\rm 2nd}$ extracted from the frequency-shift fits using the loss-derived value of $T^{\rm 2nd}_c$.  Vertical error bars indicate the spread obtained by refitting over the loss-allowed range of $T^{\rm 2nd}_c$, while horizontal error bars indicate the wafer-scale spatial variation in film stress. Both amplitudes increase strongly as the stress becomes more compressive, consistent with a greater contribution from the secondary component. The diamond marker in both panels represents the \mbox{$-542$-MPa} sample measured after cooling in a magnetic field of 300\,$\mu$T (Appendix~\ref{appendix:field_cooling}). For visual clarity, the diamond is shifted horizontally by $30$\,MPa, and its error bars, which are comparable to those of the corresponding nominally zero-field points, are omitted.
}
    \label{fig:amp_vs_stress}
\end{figure}

The stress dependence appears primarily in the amplitudes $A^\mathrm{2nd}_s$ and $B^\mathrm{2nd}_s$, indicating that the effective microwave response due to the secondary component correlates with the compressive stress, as shown in Fig.~\ref{fig:amp_vs_stress}. In contrast, under the measurement conditions, the fitted residual power-saturated TLS contribution shows no corresponding systematic dependence on film stress. Although the TLS contribution is required to reproduce the low-temperature behavior of the resonators and, for several samples, accounts for much of the initial temperature evolution, the primary evolution with stress occurs through the secondary low-gap component.

This is consistent with a scenario in which compressive stress is accompanied by an increase in the effective microwave contribution of the low-gap component rather than by a strong modification of the intrinsic transition temperature of the dominant bcc-Nb phase. Furthermore, the similar stress dependence of the dissipative and reactive amplitudes strengthens the interpretation that compressive stress increases the effective microwave contribution of the secondary low-gap component. 

Finally, Equations~\eqref{eq:minimal_loss_model} and~\eqref{eq:freq_shift} should, strictly speaking, include additional bcc-Nb quasiparticle contributions. However, we omit these terms from the final model because the data do not independently constrain the corresponding weights. Over the experimental temperature range ($100$--$900\,\mathrm{mK}$) and resonator frequencies ($4.2$--$6.3\,\mathrm{GHz}$), the Mattis--Bardeen dissipative basis function for bcc-Nb is less than $2.2\times10^{-6}$ of the corresponding secondary-component basis function, while the bcc-Nb reactive basis function is less than $1.4\times10^{-3}$ of the corresponding secondary-component basis function. The remaining reactive contribution therefore varies too weakly to be independently constrained by the data in the presence of the TLS and secondary-component contributions. For temperatures below 1\,K and the resonator frequencies used in this work, these terms have a negligible effect on the fits, and the corresponding weights cannot be reliably extracted.

\section{Discussion}

The primary result of this work is that film stress is strongly correlated with the microwave response of sputtered Nb films, whereas the transport properties vary little across the same stress range. Highly compressive films exhibit higher temperature-dependent microwave loss and larger frequency shifts, consistent with the presence of a low-gap component that is not apparent from transport measurements alone. We also emphasize that the stress values reported here are measured at room temperature, whereas film stress at cryogenic temperatures is expected to differ due to a mismatch in the thermal expansion coefficients. The film with room-temperature stress of $-371$\,MPa, which yielded the highest-quality resonators, is expected to have the smallest-magnitude stress at cryogenic temperatures among the measured films; the crude estimate in Appendix~\ref{appendix:cryogenicstress} gives a stress shift of approximately $+240$\,MPa.

The data presented here are insufficient to uniquely determine the microscopic origin, spatial distribution, or volume fraction of the low-gap regions, or the extent to which the superconducting gap in these regions is modified by proximitization effects. Possible origins of the low-gap component include an inhomogeneous-gap landscape or an additional crystalline phase. The latter possibility is consistent with the formation of the recently discovered $\omega$-phase in Nb thin films~\cite{Lee2023}. In fact, the XRD data are better described by a two-peak model (Appendix~\ref{appendix:XRD}), which may indicate the presence of an additional crystalline component. The value $T_c^\mathrm{2nd} = 2.4 \pm 0.2$\,K is consistent with the predicted value of $\sim\!2.3$\,K for $\omega$-Nb~\cite{Lee2023}. However, if $\omega$-Nb is a normal metal, the effective $T_c^\mathrm{2nd}$ could also be related to a proximity-induced gap from adjacent bcc-Nb regions. This interpretation is also consistent with the elevated power-independent loss of the most compressively stressed films if they contain sufficiently large $\omega$-Nb grains with only weakly proximitized interiors. Distinguishing among these possible microscopic origins requires additional structural characterization. Given this uncertainty, we fit the data to a~reduced version of Eq.~\eqref{eq:sigma_eff_schematic}, in which we fix the temperature dependences using the calculated response functions and absorb the unknown participation factors into amplitudes that describe films with different stress levels. Thus, the fitted amplitudes should be interpreted as effective dissipative and reactive weights, not direct measurements of the secondary-component fraction.

Additional field-cooling measurements distinguish the increased low-temperature loss from the anomalous thermal response. In particular, cooling the \mbox{$-542$-MPa} sample in 300\,$\mu$T substantially increases the low-temperature loss (Appendix~\ref{appendix:field_cooling}), consistent with additional dissipation from trapped vortices~\cite{Song2009}, while the fitted temperature-dependent dissipative and reactive amplitudes remain comparable to those obtained under a nominally zero field (Fig.~\ref{fig:amp_vs_stress}). This comparison disfavors a simple explanation in which the anomalous response scales with the trapped-vortex density, while the per-vortex response remains unchanged.

Regions with a reduced superconducting gap harbor lower-energy quasiparticle states and modify the superfluid response, potentially contributing to excess microwave loss, frequency shifts, and quasiparticle trapping. Regardless of the underlying mechanism, these results demonstrate that conventional film-quality metrics such as resistivity, residual resistance ratio, and superconducting transition temperature are insufficient predictors of microwave performance. Films with nearly identical transport properties can exhibit dramatically different microwave loss and frequency response. The presented measurements establish a correlation between stress and microwave response, showing that film stress, often treated primarily as a mechanical property, is linked to the microwave response of niobium films.

\section*{Acknowledgments}

We thank Tathagata Banerjee, Zhaslan Baraissov, Hongbin Yang, and David Muller for helpful discussions and Daniel Ralph for sharing some of the equipment used in this work. This material is based upon work supported by the Air Force Office of Scientific Research under award number FA9550-23-1-0706. Any opinions, findings, and conclusions or recommendations expressed in this material are those of the author(s) and do not necessarily reflect the views of the United States Air Force. This work was performed in part at the Cornell NanoScale Facility, a member of the National Nanotechnology Coordinated Infrastructure (NNCI), which is supported by the National Science Foundation (Grant NNCI-2025233). This work made use of the Cornell Center for Materials Research (CCMR) shared instrumentation facility.

\renewcommand{\thefigure}{S\arabic{figure}}
\renewcommand{\thetable}{S\arabic{table}}
\setcounter{figure}{0}
\setcounter{table}{0}

\appendix

\section{Fabrication}
\label{appendix:fab}
We fabricated resonators with a single-layer photolithography process as follows.
\begin{enumerate}
    \item High-resistivity 100-mm Si(100) wafers of thickness 525\,$\mu\mathrm{m}$ are treated with 2\% hydrofluoric acid (HF) for 60\,s to remove native oxides and transferred within about 5\,min to the sputter system to minimize oxide reformation. This step was omitted for the wafers used in film-stress measurements.
    
    \item The Nb film is deposited via Ar plasma sputtering in an AJA Orion 5 magnetron sputter system using the sputter parameters listed in Table~\ref{tab:sputter} and a~3-inch Nb sputter target of 99.998\% purity. The chamber base pressure before starting the process was approximately $1\times10^{-8}$\,Torr. (The last film in Table~\ref{tab:sputter} was sputtered in a different system at the Cornell Center for Materials Research (CCMR), as outlined in Ref.~\cite{Olszewski2025}, but underwent the same subsequent fabrication steps.)
    
    \item Upon removal from the deposition chamber, the wafer is prepared for photolithography by spinning AZ1512 photoresist at 4200\,rpm, which is then baked at 90°C for 60\,s.
    
    \item The pattern is written on the wafer by a Heidelberg $\mu$MLA Maskless Aligner at an~exposure dose of $160\,\mathrm{mJ}/\mathrm{cm}^2$ and a defocus setting of $-16$. The optimal defocus value must be determined by dose testing, as it is tool-dependent.
    
    \item We develop the wafer in AZ726 MIF developer for 60\,s, rinse with deionized (DI) water with resistivity above $1\,\textrm{M}\Omega\!\cdot\!\textrm{cm}$, and hard-bake at 110°C to ensure feature hardness for reactive ion etching. There was a delay of approximately 24\,h before the subsequent processing steps at the Cornell NanoScale Facility (CNF).
    
    \item At CNF, the wafer is descummed in an oxygen plasma for 2\,min using an Anatech plasma asher with 220\,sccm of O\textsubscript{2} and 20\,sccm of N\textsubscript{2} at 150\,W.
    
    \item The wafer is then etched in a Plasma-Therm 770 inductively coupled plasma (ICP) etch system using a two-stage recipe. First, the sample is subjected to a light etch with a mixture of BCl\textsubscript{3}, Cl\textsubscript{2}, and Ar gas in a 2:30:5 volumetric flow ratio with reactive-ion etch (RIE) and ICP powers of 26\,W and 800\,W at 13\,mTorr. This is followed by a primary etch with a mixture of BCl\textsubscript{3}, Cl\textsubscript{2}, and Ar gas in a 30:20:5 volumetric flow ratio with RIE and ICP powers of 12\,W and 800\,W at 7\,mTorr.
    
    \item After etching, we strip the photoresist in successive ``dirty'' and ``clean'' baths of AZ300T heated to 80°C for 10\,min in each bath, then thoroughly rinse in DI water. Because AZ300T contains tetramethylammonium hydroxide (TMAH), this step also helps remove chlorine-based etch byproducts~\cite{Olszewski2025}.
    
    \item The wafer is descummed in an oxygen plasma for 2\,min using an Anatech plasma asher with 220\,sccm of O\textsubscript{2} and 20\,sccm of N\textsubscript{2} at 150\,W.
    
    \item Another layer of S1813 photoresist is spun as a protective layer for dicing and baked at 90°C for 60\,s.
    
    \item We dice the wafer using a DISCO dicing saw.
    
    \item Before wire bonding for measurements, the protective photoresist is stripped by sonicating each sample in acetone for 5\,min, followed by an isopropyl alcohol (IPA) rinse.
    
    \item The samples are affixed to the Al sample packages using GE varnish and then wire-bonded. Additional Al ground-plane stitching bonds are made across the resonators and transmission line.
\end{enumerate}
Neither buffered oxide etching nor hydrofluoric-acid treatment was performed after step~12. These treatments were deliberately omitted to avoid introducing Nb hydrides that could confound the measured high-power loss~\cite{McRae2020, Torres-Castanedo2024, Olszewski2025}.

We tested 11 sputter recipes, as summarized in Table~\ref{tab:sputter}. The measured film stress generally exhibits the expected dependence~\cite{Windischmann1991} on the Ar gas pressure and the throw distance (the distance between the Nb target and the Si wafer). We selected five sputter recipes to fabricate resonators, as highlighted in the table.

\begin{table}[ht]
    \centering
    \begin{tabular}{l l l l} 
 \hline\hline
 Ar pressure & Throw distance & Film thickness & Film stress\\
 (mTorr) & (mm) & (nm) & (MPa)\\ 
 \hline
  2 & 30 & 65 & $-982\pm82$ \\ 
 \textbf{2} & \textbf{70} & \textbf{65} & $\mathbf{-737\pm72}$ \\ 
 \textbf{6} & \textbf{30} & \textbf{65} & $\mathbf{-542\pm56}$ \\ 
 6 & 30 & 111 & $-598\pm41$ \\ 
 6 & 30 & 228 & $-527\pm33$ \\ 
 \textbf{6} & \textbf{70} & \textbf{65} & $\mathbf{-371\pm33}$ \\ 
 6 & 70 & 111 & $-373\pm26$ \\ 
 6 & 70 & 228 & $-400\pm22$ \\ 
 10 & 30 & 65 & $-408\pm60$ \\ 
 \textbf{10} & \textbf{70} & \textbf{65} & $\mathbf{-56\pm 47}$ \\
 \textbf{2} & \textbf{250} & \textbf{60} & $\mathbf{\phantom{-}100\pm77}$ \\
\hline\hline
\end{tabular}
\caption{Sputter parameters and corresponding Nb film stress. Films used to fabricate the resonators are highlighted in \textbf{bold}. All films were sputtered at an Ar gas flow rate of 60\,sccm except the one with the 250\,mm throw distance, which was sputtered at 70\,sccm.}
\label{tab:sputter}
\end{table}

\section{Profilometry}
\label{appendix:profilometery}

\begin{figure}
    \centering
    \includegraphics[width=\linewidth]{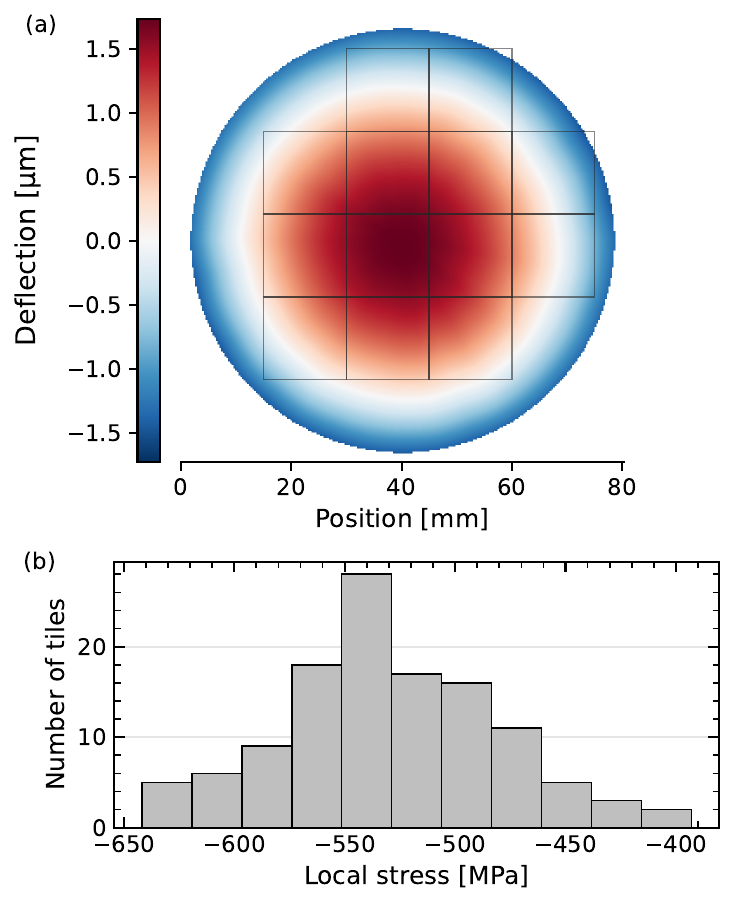}
    \caption{\textbf{Estimation of the spatial variation of the film stress.} (a)~Measured wafer deflection map obtained by stylus profilometry. For visualization, the wafer is overlaid with representative non-overlapping $15 \times 15\,\mathrm{mm}^2$ regions; the analysis was performed using overlapping regions of the same size with a 5\,mm center-to-center spacing. Each region was independently fit to a quadratic surface to estimate the local wafer curvature, which was then converted to a local film stress using the Stoney relation. (b)~Distribution of the resulting local stress estimates. The mean of the local stress estimates reproduces the stress obtained from the full-wafer fit, while the standard deviation indicates the wafer-scale spatial variation in film stress.}
    \label{fig:stress_wafer_tile}
\end{figure}

We used a separate set of wafers for profilometer measurements, with the films deposited under the same sputter conditions as those used to fabricate the CPW resonators. The wafers used for the stress measurements did not receive any pre-sputter surface treatment (step~1, Appendix~\ref{appendix:fab}).

Film stress was determined from wafer bow measurements obtained using a KLA P16+ stylus profiler. We performed 18 scans across the wafer diameter, each separated by a $10^\circ$ rotation, from which the instrument reconstructed a two-dimensional wafer topography. The measurement was performed on each wafer both before and after Nb deposition, allowing us to determine the change in wafer deflection due to the deposited film.

We independently analyzed the exported difference map, $z(x,y)$, to determine wafer curvature. The measured height difference was fit to a radially symmetric quadratic surface on a tilted plane
\begin{align}
z(x,y)=&
A\left[(x-x_0)^2+(y-y_0)^2\right] \nonumber\\
&+B(x-x_0)+C(y-y_0)+D,
\end{align}
where $(x_0,y_0)$ denotes the center of the wafer and $A$, $B$, $C$, and $D$ are fit coefficients. The linear terms account for small residual wafer tilt, while the quadratic term describes the wafer bow. For a spherical surface, the wafer curvature is $\kappa=1/R=2A$. The corresponding biaxial film stress is calculated using the Stoney relation~\cite{Freund2004}:
\begin{equation}
\sigma_f=\frac{E_s t_s^2}{6(1-\nu_s)t_f}\,\kappa,
\label{eq:Stoney}
\end{equation}
where $E_s$, $\nu_s$, and $t_s$ are Young's modulus, Poisson's ratio, and the thickness of the Si substrate, respectively, and $t_f$ is the measured thickness of the Nb film. A substrate biaxial modulus of $E_s/(1-\nu_s)=180.3\,\mathrm{GPa}$~\cite{JANSSEN20091858}, corresponding to Si(100), and a substrate thickness of $525\,\mu\mathrm{m}$ are used. The stresses calculated from the independently fitted curvatures are in excellent agreement with the values reported by the commercial profilometer software, thus cross-validating the analysis.

To estimate the spatial variation in film stress across each wafer, the wafer was divided into overlapping $15\times15\,\mathrm{mm}^2$ regions with a 5-mm center-to-center spacing. The same quadratic surface fit was applied independently to each region to obtain a local curvature, which was converted to a local film stress using the Stoney relation~\eqref{eq:Stoney} (Fig.~\ref{fig:stress_wafer_tile}). The mean of the local stress estimates agrees with the stress obtained from the full-wafer fit, demonstrating that the local fitting procedure accurately reproduces the global wafer curvature. The standard deviation of the local stress estimates is taken as a measure of spatial variation in film stress at the wafer scale and is reported in Table~\ref{tab:sputter}. Because these measurements were performed on intact wafers prior to dicing and device fabrication, the reported values should be interpreted as the as-deposited wafer-level film stress and its spatial variation rather than the exact residual stress of an individual fabricated device.

\section{X-ray diffraction (XRD)}
\label{appendix:XRD}

XRD measurements were performed on the first 10 wafers listed in Table~\ref{tab:sputter}. These were the same wafers characterized by profilometry in Appendix~\ref{appendix:profilometery} and were prepared as a companion set to the wafers used to fabricate the resonator samples.

XRD measurements were performed at CCMR using a~Rigaku SmartLab X-ray diffractometer equipped with a~Cu $K\alpha$ source and a~Ge(220) 2-bounce monochromator. Scans were acquired with an angular step size of $0.02^\circ$, with the attenuator in automatic mode and with the software-controlled incident and receiving slits set to $1$\,mm. The manual slit that controls the width of the X-ray beam was set to $15$\,mm.

\begin{figure}
    \centering
    \includegraphics[width=\linewidth]{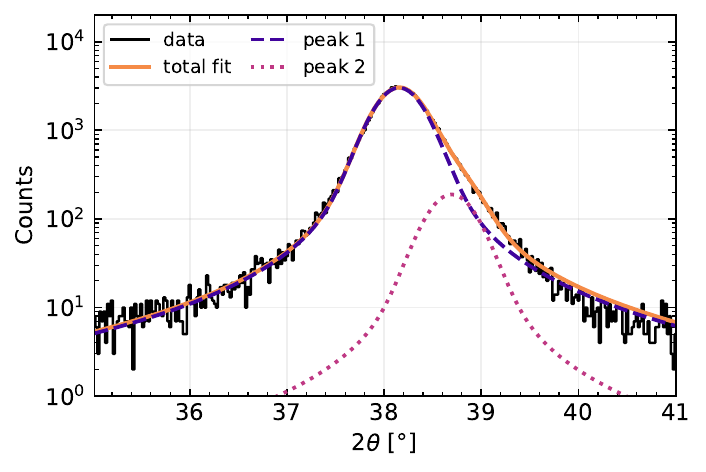}
    \caption{\textbf{XRD analysis.} Typical XRD fit showing both the primary bcc-Nb peak and the small secondary peak to the right, which may originate from a secondary crystalline component such as $\omega$-Nb. Data shown for the film with an~intrinsic stress of $-737$\,MPa.}
    \label{fig:FigApp_Profile}
\end{figure} 

XRD scans were analyzed using a Poisson maximum-likelihood framework. The number of counts $k_i$ at the diffraction angle $2\theta_i$ is modeled as an independent Poisson random variable, with the probability of observing exactly $k_i$ counts given by
\begin{equation}
P(k_i|\lambda_i)=\frac{\lambda_i^{k_i}e^{-\lambda_i}}{k_i!},
\end{equation}
where $\lambda_i$ is the expected number of counts predicted by the diffraction model. The likelihood of a complete scan is therefore
\begin{equation}
\mathcal{L}=\prod_i P(k_i|\lambda_i).
\end{equation}
Maximizing the likelihood is equivalent to minimizing the Poisson negative log-likelihood
\begin{equation}
-\ln\mathcal{L}=\sum_i
\left[\lambda_i-k_i\ln\lambda_i\right]
+\mathrm{const},
\end{equation}
where the constant absorbs the terms independent of the diffraction model choice.

The expected diffraction intensity was modeled as a~constant background~$\lambda_0$ together with one or two Voigt diffraction peaks~\cite{Cullity2001, Thompson1987},
\begin{equation}
\lambda(2\theta)=\lambda_0 +
\sum_{p=1}^{N_p} A_p V(2\theta;\mu_p,\upsilon,\gamma),
\end{equation}
where $V$ denotes a Voigt profile with center $\mu_p$, Gaussian width $\upsilon$, and Lorentzian width $\gamma$. $N_p$ is the number of peaks in the model, either one or two. For the two-peak hypothesis, the Gaussian and Lorentzian widths were constrained to be identical for both peaks, while the amplitudes were fit independently. This reflects the expectation that the two components are measured under identical instrumental conditions and share comparable broadening mechanisms, while avoiding unnecessary parameter degeneracies that are not supported by the available counting statistics.

To avoid convergence to local minima, we initialized the two-peak model from multiple starting values for the second-peak amplitude and peak separation and retained the fit with the lowest negative log-likelihood.

Model selection between the one- and two-peak hypotheses was performed using the Akaike Information Criterion (AIC)~\cite{AIC, Burnham2002} defined as
\begin{equation}
\mathrm{AIC}=2q-2\ln\mathcal{L},
\end{equation}
where $q$ is the number of free parameters: five for the one-peak model and seven for the two-peak model. We use the difference between the AIC values for two- and one-peak models
\begin{equation}
\Delta\mathrm{AIC}=\mathrm{AIC}(N_p=2)-\mathrm{AIC}(N_p=1)
\end{equation}
to compare the two, with negative values indicating preference for the two-peak description. For each analyzed diffraction scan, the two-peak model was favored over the corresponding single-peak model based on a negative $\Delta$AIC, providing some additional evidence for a second diffraction component. Although the fitted amplitudes may be interpreted as estimates of the relative scattering contributions from the two diffraction components, the limited number of diffraction scans and parameter uncertainties preclude a robust determination of the stress dependence of the secondary component fraction. We therefore rely primarily on the XRD data as evidence for an additional diffraction component, while the quantitative dependence of the additional component on film stress remains a subject for future investigation with a larger diffraction dataset.

To confirm that the XRD peak position correlates with film stress, we compared the profilometry and XRD measurements. The linear fit to the data in Fig.~\ref{fig:Fig1}(b) gives a~slope of $0.50\pm0.04^\circ/\textrm{GPa}$. For an isotropic film under equibiaxial plane stress, the out-of-plane normal strain is \mbox{$\epsilon_{zz}=-2\nu_f\sigma/E_f$}. Differentiating Bragg's law gives $d(2\theta)=-2\tan\theta\,d\epsilon_{zz}$, and therefore $d(2\theta)/d\sigma = 4\nu_f\tan\theta/E_f$, where $\theta$ is the Bragg angle~~\cite{Cullity2001}. Using Young's modulus~$E_f=104.9\,\textrm{GPa}$ and Poisson's ratio $\nu_f=0.397$ for bulk Nb~\cite{Emsley1997}, we obtain $4\nu_f\tan\theta/E_f\approx0.30^\circ/\textrm{GPa}$ for $\theta=19.1^\circ$, whereas the fitted slope is approximately 1.7 times this estimate. This discrepancy likely reflects the limitations of applying a~simple isotropic bulk elastic model to textured sputtered Nb films as well as possible deposition-dependent changes in the stress-free lattice parameter. However, our independent analysis of exported profilometry data confirms that the instrument-reported stress values are consistent with the values inferred from the wafer curvature (Appendix~\ref{appendix:profilometery}), indicating that the discrepancy is unlikely to arise from the curvature-to-stress conversion implemented by the profilometer software.

\section{Microwave measurements}
\label{appendix:setup}

\begin{figure}
    \centering
    \includegraphics[width=.5\linewidth]{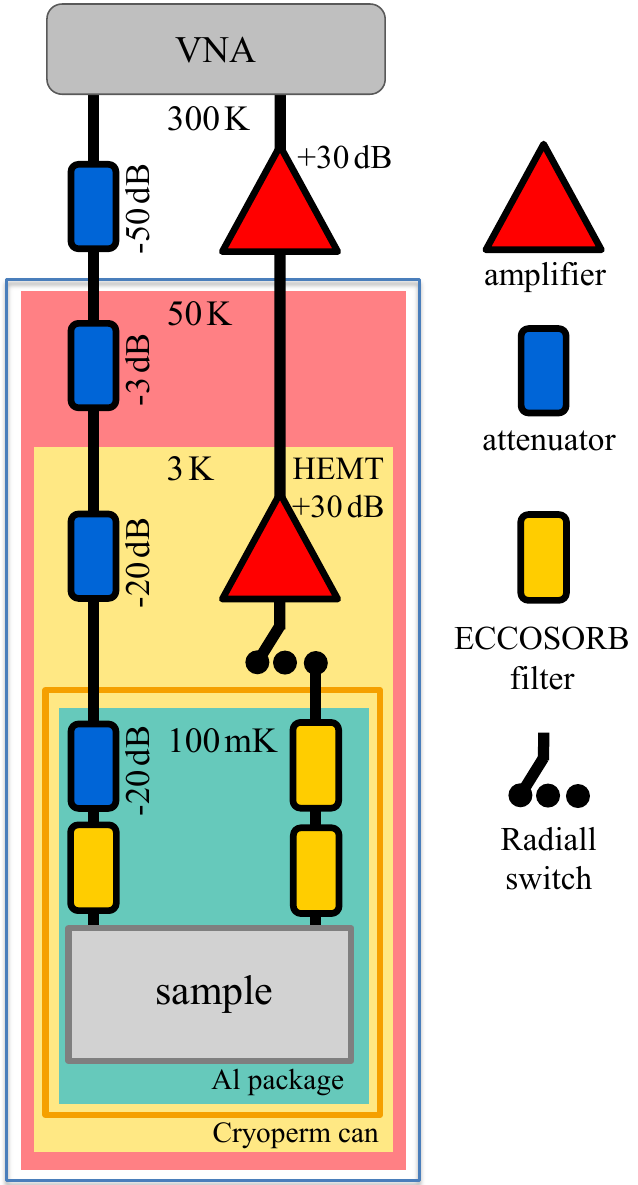}
    \caption{\textbf{Microwave measurement setup.} Schematic of the cryogenic microwave measurement setup used to characterize the CPW resonators. The resonator chip was mounted in an Al package on the 100\,mK stage of an~adiabatic demagnetization refrigerator. The 100\,mK stage temperature could be swept from 70\,mK to 3\,K. The sample package was placed inside a Cryoperm magnetic shield anchored to the 3\,K stage.}
    \label{fig:FridgeDiagram}
\end{figure} 

We measured the CPW resonator samples in a High Precision Devices (HPD) Model 106 Shasta adiabatic demagnetization refrigerator (ADR). Figure~\ref{fig:FridgeDiagram} shows the measurement setup. The packages were fastened to an~oxygen-free high-conductivity (OFHC) copper cold finger cooled to a minimum temperature of 70\,mK. A Cryoperm shield anchored to the 3\,K stage surrounded the sample, along with three vacuum cans; the inner two cans were covered with a black infrared-absorbing coating to minimize blackbody radiation. 

The input line included 50\,dB of room-temperature attenuation, 43\,dB of added cryogenic attenuation, and an~ECCOSORB filter thermalized to the cold finger. The output line included ECCOSORB filters, a~Caltech CITCRYO4-12A high-electron-mobility transistor (HEMT) amplifier at the 3\,K stage and a Narda Microwave-West DBS-0208N315 room-temperature amplifier. All $S_{21}$ measurements were performed with an Agilent N5230A vector network analyzer. The temperature was measured with a RuOx thermometer mounted on the same stage as the CPW resonator samples.
 
After completion of the primary measurements, an additional resonator chip from the $-371$\,MPa wafer was measured without the aluminum stitching bonds that connect the ground-plane patches across the coplanar waveguide resonator. Fitting the temperature-dependent loss and frequency shift with $T_c^{\rm 2nd}$ fixed to the value extracted from the primary sample series yielded dissipative and reactive amplitudes consistent with the stress-dependent trends reported in the main text. This agreement indicates that the stitching bonds are not responsible for the anomalous microwave response.
 
The stitching bonds were originally included to suppress possible slotline modes. However, aluminum stitching bonds become normal near $T_c^{\rm Al}\approx1.2$\,K, introducing an additional loss channel that limits the usable temperature range of the resonator measurements. We found no evidence that such modes limited measurements on an unstitched device. Thus, omitting these bonds in future measurements could allow the intrinsic film response to be followed to higher temperatures, provided slotline modes remain well outside the measurement band.

\section{Photon number during temperature measurements}
\label{appendix:photon_number}

\begin{figure}
    \centering
    \includegraphics[width=.8\linewidth]{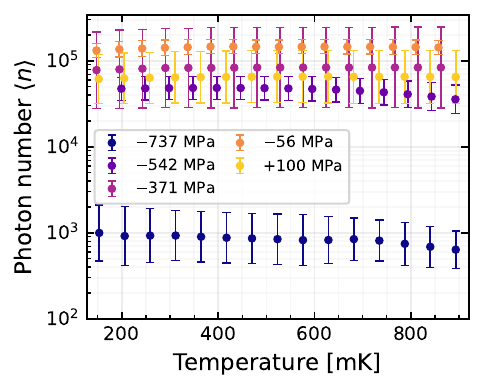}
    \caption{\textbf{Photon number during temperature sweeps.} Estimated photon number $\langle n \rangle$ as a function of temperature~$T$. Over the fitting temperature range, the photon number remains sufficiently high and changes little, so the TLS saturation factor varies only slightly.}
    \label{fig:FigApp_TempPhotonNumber}
\end{figure} 

\begin{figure*}
    \centering
    \includegraphics[width=\linewidth]{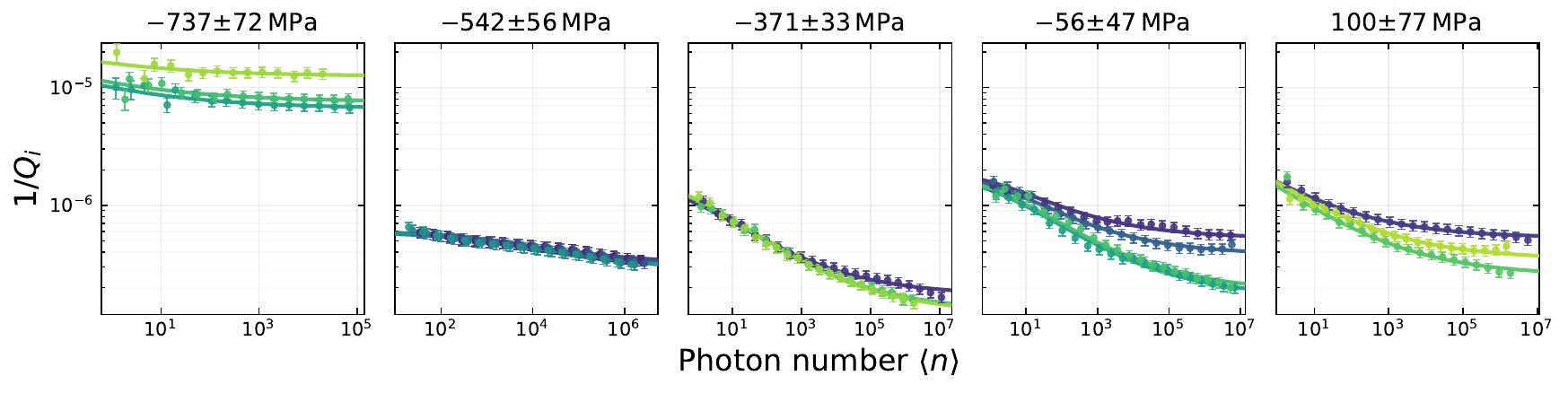}
    \caption{\textbf{Power-dependent microwave loss.} Inverse internal quality factors $1/Q_i$ as a function of resonator photon number~$\langle n \rangle$ for samples of various film stresses. For each sample, solid lines show simultaneous fits to Eq.~\eqref{eq:power_model} with a shared TLS loss amplitude and a characteristic saturation photon number, but with a resonator-specific background loss.}
    \label{fig:power_fits}
\end{figure*} 

The temperature-dependent measurements discussed in the main text were performed at fixed microwave drive power. Because the resonator photon population $\langle n \rangle$ depends on the resonator quality factor, the number of photons can decrease with increasing temperature as the resonators become more lossy. The photon number for each resonator was therefore calculated at all temperatures using Eq.~(3.33) from Ref.~\cite{deVisser2014}:
\begin{equation}
    \langle n \rangle  = \frac{2P}{\hbar\omega^2} \frac{Q^2}{Q_c},
\end{equation}
where $\omega$ is the angular frequency of the resonator, $Q=(Q_i^{-1}+Q_c^{-1})^{-1}$ is the total quality factor, $Q_i$ and $Q_c$ are the internal and external quality factors, and $P$ is the total power at the sample level.

Figure~\ref{fig:FigApp_TempPhotonNumber} shows the calculated photon number $\langle n \rangle$ throughout the temperature sweep. The photon number remains sufficiently high that the TLS saturation factor changes only slightly over the corresponding temperature range (Appendix~\ref{appendix:power}). Consequently, the resonant TLS contribution is expected to remain mostly saturated.

\begin{figure}
    \centering
    \includegraphics[width=.8\linewidth]{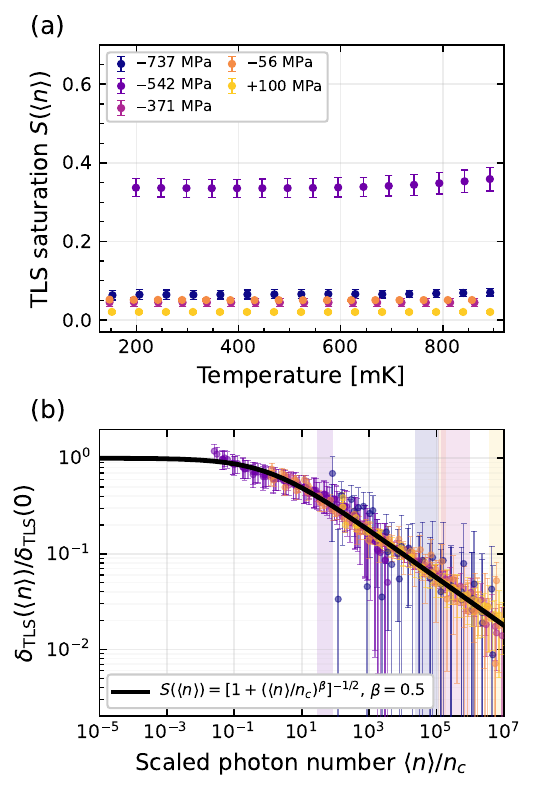}
    \caption{
    \textbf{TLS saturation during the temperature sweeps.}
    (a)~Average TLS saturation factor $S(\langle n\rangle)=[1+(\langle n\rangle/n_c)^\beta]^{-1/2},\,\beta=0.5,$ for each film in the temperature-dependent measurements. Error bars indicate the resonator-to-resonator standard deviation within each sample. Over the presented temperature range, the saturation factor remains nearly constant despite changes in resonator frequency and quality factor, indicating that the TLS power-saturation factor can be treated as approximately temperature-independent during the temperature sweeps. (b)~Collapse of the normalized TLS loss measured in the power-dependent experiments onto the universal saturation functional form of Eq.~\eqref{eq:collapse}. Shaded bands indicate the ranges of normalized photon numbers $\langle n\rangle/n_c$ sampled for each film over the temperatures used in the fit to the model. All temperature sweeps remain within the TLS-saturated regime, where the TLS contribution is reduced to approximately 5--35\% of its unsaturated value.}
    \label{fig:power_collapse}
\end{figure} 

\begin{figure}
    \centering
    \includegraphics[width=.8\linewidth]{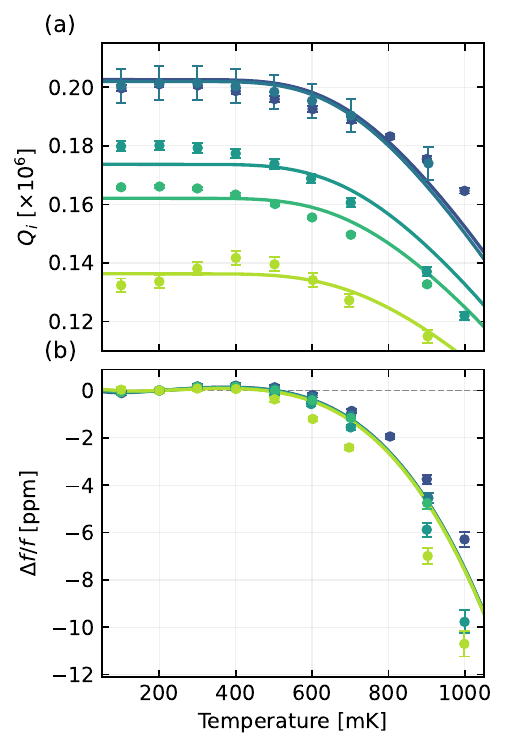}
    \caption{\textbf{Temperature-dependent microwave response after magnetic field cooling.} (a)~Internal quality factors~$Q_i$ and (b)~frequency shifts $\Delta f/f$, referenced to 100\,mK, for resonators fabricated from the $-542$\,MPa film after cooling in 300\,$\mu$T. Solid curves show the fit with the secondary component transition temperature fixed at $T_c^{\rm 2nd}=2.4\,$K, the value obtained from the nominally zero-field stress-series analysis. Resonators are colored by their resonance frequency, with lighter colors representing higher frequencies.}
    \label{fig:high_field}
\end{figure}

\section{Power-dependent TLS characterization}
\label{appendix:power}

Given the high drive power used, resonant TLS loss is expected to be substantially saturated. To verify this assumption, we independently measured the internal quality factor as a function of the photon number~$\langle n \rangle$ at the reference temperature of $T_{\rm ref} = 100$\,mK.

At a fixed temperature, the microwave loss for resonator~$j$ from sample~$s$ is modeled as
\begin{equation}
\frac{1}{Q_{i;s,j}(\langle n\rangle)}=\delta'_{0;s,j}+
\frac{\delta^{\rm TLS}_s}{\sqrt{1+\left(\langle n\rangle/n_{c;s}\right)^\beta}},
\label{eq:power_model}
\end{equation}
where $\delta'_{0;s,j}$ is a photon-independent background loss, $\delta^{\rm TLS}_s$ is the unsaturated TLS loss amplitude at a fixed low temperature, $n_{c;s}$ is the characteristic saturation photon number, and $\beta$ is the TLS interaction exponent~\cite{Gao2008, McRae2020}. For simplicity, we fix the TLS interaction exponent to $\beta=0.5$. For each sample, the unsaturated TLS loss amplitude and characteristic saturation photon number are shared across all resonators, while the background loss is allowed to vary independently for each resonator. This parameterization reflects the expectation that resonators fabricated from the same film share a~similar TLS environment but differ in resonator-specific background losses.

Figure~\ref{fig:power_fits} shows the resulting fits for all resonators. The model describes the observed power dependence across the measured stress range. The fitted characteristic saturation photon numbers $n_{c;s}$ vary substantially between samples but remain below the photon numbers used in the temperature sweeps. This confirms that the temperature-dependent measurements presented in the main text were performed within the TLS-saturated regime.

To further test the model's consistency, we subtract the resonator-specific background loss and normalize each measurement by its fitted TLS loss amplitude. After additionally scaling the photon number by the fitted characteristic saturation photon number, all resonator data collapse onto the universal TLS saturation curve given by
\begin{equation}
\frac{1/Q_{i;s,j}(\langle n\rangle)-\delta'_{0;s,j}}
{\delta^{\rm TLS}_s}=\frac{1}
{\sqrt{1+\left(\langle n\rangle/n_{c;s}\right)^\beta}}.
\label{eq:collapse}
\end{equation}
The resulting collapse is shown in Fig.~\ref{fig:power_collapse}. The agreement demonstrates that, once differences in resonator background loss, TLS loss amplitude, and characteristic saturation photon number are accounted for, the power dependence follows the conventional TLS saturation model over several decades in normalized photon numbers. The shaded regions in Fig.~\ref{fig:power_collapse} denote the approximate range of normalized photon numbers sampled during the temperature-dependent measurements. Throughout this range, the TLS contribution is suppressed to approximately 5--35\% of its unsaturated value.

\section{Mattis--Bardeen basis functions}
\label{appendix:MB}

The approximate Mattis--Bardeen basis functions are
\begin{align}
\frac{\sigma_1(f,T;T_c)}{\sigma_n}&=
\frac{4\Delta(T)}{hf}e^{-\frac{\Delta(T)}{k_BT}}
\sinh\!\left(x\right)K_0\!\left(x\right),\label{eq:sigma1_mb_approx}
\\
\frac{\sigma_2(f,T;T_c)}{\sigma_n}&=
\frac{\pi\Delta(T)}{hf}\left[1-2e^{-\frac{\Delta(T)}{k_BT}}e^{-x}I_0(x)\right],\label{eq:sigma2_mb_approx}
\end{align}
where $x \equiv hf/(2k_BT)$,
$I_0(x)$ and $K_0(x)$ are the modified zero-order Bessel functions of the first and second kinds, $\Delta(T)$ is the superconducting energy gap at temperature $T$, and $\sigma_n$ is the normal-state conductivity used to normalize the complex conductivity~\cite{Gao2008a}. We assume the Bardeen--Cooper--Schrieffer (BCS) relation $\Delta(0)=1.764\,k_B T_c$ and the standard BCS temperature dependence of $\Delta(T)$.

Equations~\eqref{eq:sigma1_mb_approx} and \eqref{eq:sigma2_mb_approx} are the low-temperature, local-limit Mattis--Bardeen forms used as temperature-dependent basis functions in the fit, rather than a~full numerical evaluation of the Mattis--Bardeen integrals~\cite{Mattis-Bardeen1958, Gao2008}. This approximation captures the activated quasiparticle dissipation through $\sigma_1$ and the corresponding reduction in superfluid density via $\sigma_2$. Since the analysis relies primarily on the temperature dependence of these functions, the unknown electrodynamic participation and kinetic-inductance factors are absorbed into the fitted amplitudes $A^{\rm 2nd}_s$ and $B^{\rm 2nd}_s$. The bcc-Nb contribution evaluated with our average measured film value of $T_c^{\rm bcc}=9.0$\,K is exponentially suppressed over the measurement range; the fitted stress-dependent quasiparticle response is therefore dominated by the secondary component.

\section{Vortex-induced loss}
\label{appendix:field_cooling}

Vortex motion can produce microwave dissipation in superconducting resonators~\cite{Song2009} and has recently been proposed as the origin of anomalous loss in Ta devices~\cite{Bahrami2025}. To test whether trapped vortices could similarly account for the temperature-dependent loss observed in our Nb resonators, we performed field-cooling measurements on a resonator chip fabricated from the $-542$\,MPa film.

We carried out the measurements in the ADR, which was equipped with a Helmholtz coil inside the Cryoperm shield. This allowed us to apply a controlled static magnetic field perpendicular to the film plane during sample cooldown. Before the measurement, the refrigerator cold finger was thermally isolated from the 3\,K stage and heated above 10\,K using a~dc heater, well above the superconducting transition temperature of the Nb film. The desired magnetic field was then applied, and the cold finger was thermally reconnected to the 3\,K stage and cooled through $T_c$ prior to the microwave response measurements.

After field cooling in 300\,$\mu$T, we performed a~temperature-dependent measurement at a~drive power corresponding to $\langle n \rangle \gtrsim 2\times10^{3}$ photons in each measured resonator to test how vortices modify the anomalous thermal loss (Fig.~\ref{fig:high_field}). Compared with the nominally zero-field measurements in Fig.~\ref{fig:Qidf_Model}, field cooling reduces the low-temperature internal quality factors by approximately an order of magnitude and produces redshifts of the resonant frequencies in the range of $750$--$880$\,ppm (not shown). These changes are consistent with additional dissipation from trapped vortices within the film and CPW center conductor. Nevertheless, fits with $T_c^{\rm 2nd}=2.4$\,K, fixed to the value obtained from the nominally zero-field analysis, yield effective dissipative and reactive amplitudes $A^{\rm 2nd}_s$ and $B^{\rm 2nd}_s$ of comparable magnitude under the two cooling conditions (Fig.~\ref{fig:amp_vs_stress}). The similarity in $B^{\rm 2nd}_s$ is directly evident from the comparison of Figs.~\ref{fig:Qidf_Model} and~\ref{fig:high_field}(b), which show that field cooling does not appreciably modify the pronounced temperature-dependent component of the redshifts observed in the compressively stressed film. Within the phenomenological model, the frequency-shift behavior and the increased loss in the magnetic field are consistent with a predominantly temperature-independent increase in inductance and an additional vortex-induced loss contribution, the latter captured by an increase in the temperature-independent loss constant~$\delta_0$.

Thus, we find no evidence that the temperature-dependent part of the response should be attributed to vortices in nominally zero and 300\,$\mu$T fields. Our results rule out a simple model in which the anomalous thermal response scales with the trapped-vortex density while the contribution from each vortex remains unchanged. We therefore conclude that trapped vortices or vortex motion are unlikely to cause anomalous thermal loss under our experimental conditions.

\section{Change in film stress upon cooling to cryogenic temperatures}
\label{appendix:cryogenicstress}

It is instructive to estimate the change in film stress as the films are cooled to cryogenic temperatures, especially given the possibility of phononic poisoning due to stress relief~\cite{Anthony-Petersen2024, Yelton25}. The change in film stress can be roughly estimated using Eq.~(26) from Ref.~\cite{Abadias2018}:
\begin{equation}
    \Delta\sigma(T) = (\alpha_s - \alpha_f)(T-T_i)\frac{E_f}{1-\nu_f},
\end{equation}
where $\alpha_s$ and $\alpha_f$ are the thermal expansion coefficients of the substrate and the film, respectively; $E_f$ and $\nu_f$ are Young's modulus and Poisson's ratio of the film, respectively; $T_i$ is the initial temperature, and $T$ is the temperature of interest.

For a Nb film on a Si substrate, the Nb elastic parameters from Ref.~\cite{Emsley1997} and the room-temperature thermal expansion coefficients from the Wolfram Language \texttt{ElementData} function~\cite{WolframElementData} ($\alpha_s = 2.6\times10^{-6}\,\textrm{K}^{-1}$, $\alpha_f=7.3\times10^{-6}\,\textrm{K}^{-1}$), together with $T_i = 293$\,K and $T \approx 0$\,K, give $\Delta\sigma=+240$\,MPa. This estimate treats the thermal expansion coefficients and elastic constants as temperature-independent and provides only an approximate change in stress.

\clearpage
\bibliography{bibgliography}

\end{document}